\documentclass[aps,prb,twocolumn,floats]{revtex4-2}
\usepackage[margin=0.8in,footskip=0.25in,paperwidth=8in,paperheight=13.5in]{geometry}
\usepackage{latexsym}
\usepackage{balance}
\usepackage{amsmath}

\usepackage{amssymb}
\usepackage{bm}
\usepackage{wasysym}
\usepackage[dvips]{color}
\usepackage{graphicx}
\usepackage{subfigure}
\usepackage{epsfig}
\usepackage{balance}
\usepackage{subfigure}
\usepackage{ulem}

\DeclareMathAlphabet{\mathpzc}{OT1}{pzc}{m}{it}

\newcommand{\hide}[1]{}
\newcommand{\veps}{\varepsilon}

\def\bfr{{\bf r}}
\def\ra{\rangle}

\def\veps{\varepsilon}

\usepackage{tikz}
\usetikzlibrary{matrix}
\usetikzlibrary{arrows}

\graphicspath{{./figures/}}
\usepackage{float}

\usepackage[colorlinks,bookmarks=true,citecolor=blue,linkcolor=blue,urlcolor=blue, breaklinks=true]{hyperref}
\usepackage[sort&compress]{natbib}

\begin{document}
\title{Chiral tunneling  in single layer graphene with Rashba spin-orbit coupling: spin currents}
\author{Y. Avishai$^{1}$ and Y. B. Band$^{2}$}
\affiliation{$^1$Department of Physics,
  Ben-Gurion University of the Negev,
  Beer-Sheva, Israel, \\
  New York University and the NYU-ECNU Institute
  of Physics at NYU Shanghai, 3663 Zhongshan Road North,
  Shanghai, 200062, China, \\
  and Yukawa Institute for Theoretical Physics, Kyoto, Japan\\
  $^2$Department of Physics, Department of Chemistry,
  Department of Electro-Optics, and
  The Ilse Katz Center for Nano-Science,
  Ben-Gurion University of the Negev,
}


\begin{abstract}
We study forward scattering of 2D massless Dirac electrons at Fermi energy $\veps>0$ in single layer graphene through a 1D rectangular barrier of height $u_0$ in the presence of uniform Rashba spin-orbit coupling (of strength $\lambda$).  The role of the Klein paradox in graphene spintronics is thereby exposed. It is shown that: (1) For $\veps-2 \lambda <u_0<\veps+2 \lambda$ there is partial Klein tunneling, the transmission coefficient $T(\lambda)<1$ and, quite remarkably, $T(\lambda \ge 0.1 \ \text{meV}) \approx 0$ when the scattering energy equals the barrier height, $\veps=u_0$, [whereas $T(\lambda=0) = 2$]. (2) Spin density and spin-current density are remarkably different than in bulk single layer graphene. In particular, they are sensitive to $\lambda$ and $u_0$.  (3) Spin current densities are space dependent, implying the occurrence of non-zero spin torque density.  Such a system may serve as a graphene based spintronic device without the use of an external magnetic field or magnetic materials.
\end{abstract}

\maketitle

\section{Introduction} \label{sec:intro}

Shortly after the discovery of graphene \cite{Geim}, numerous novel physical phenomena were revealed in its electronic properties \cite{Guinea,Sarma}. One of these, the occurrence of chiral tunneling and the Klein paradox \cite{Klein} in single layer graphene (SLG), was reported in a seminal paper \cite{Katsnelson_06}.  This work was further expanded in Refs.~\cite{Cheianov, Katsnelson-2012, AF_11}.  It was shown that, due to chirality near a Dirac point, electrons execute unimpeded transmission for energies below the potential barrier. This phenomenon is related to the absence of back-scattering for electron-impurity scattering in carbon nanotubes \cite{Ando}.  Several additional extensions have been reported in Refs.~\cite{Peeters, Barbier,AB}.  In parallel, investigation of the role of electron spin in graphene led to the emergence of a new field: graphene spintronics \cite{Huertas, Min,Yao,Castro, Trau,Tombros, Tombros1,Cho, Zarea, Rashba1,Liu, Gmitra0,  Berc,Sergej1,Schwierz,Tse,Sergej2, Miao, Jo, Liu2, Mar, Richter, Patra, Dulbak, Zomer, Zhang1,Lenz, Shakouri, Tang, Jaros, Kochan, Tuan, Gmitra, Ferrari, Roch, Zhang, Drogler, Avsar, Ingla, Medina, Li, Lin, Amir, Shifei}.

Here we study Klein tunneling of 2D massless Dirac electrons at Fermi energy $\veps>0$ in SLG, through a 1D barrier of height $u_0$ in the presence of uniform RSOC of strength $\lambda$ \cite{Rashba}. The motivation is to elucidate the effect of the Klein paradox on observables such as spin density, spin current density and spin torque \cite{Niu}.  The calculated  spin observables have properties that are remarkably different from those predicted in bulk SLG \cite{Zhang} (i.e., in the absence of a 1D potential so that the Klein paradox does not play a role).  In particular, symmetry relations are broken (see below), and the spin current is space dependent so that there is a finite spin torque. But most importantly, the response of the spin densities to the RSOC strength is substantial even for small RSOC coupling (the strength of Rashba splitting caused by a strong electric field in SLG  is reported in Ref.~\cite{Gmitra} to be a fraction of an meV). Hence, we hope that our study will motivate the fabrication of graphene based spintronic devices that do not rely on the use of an external magnetic field or magnetic materials.

In addition to the study of spin related observables,
we also expose the occurrence of {\it partial Klein tunneling} which causes an intriguing  behaviour of the transmission coefficient $T$.  
Explicitly, the pertinent scattering problem involves two scattering channels. In the narrow region $\veps-2 \lambda<u_0<\veps+2 \lambda$, (for $\lambda > 1$ meV), only one scattering channel is open, and $T<1$.  Moreover, for scattering energy equal to the barrier height, $\veps=u_0$, one channel just closes while the second channel just opens, and $T \approx 0$ (recall that $T=2$ for $\lambda=0$).

In the sections below we present the formalism used (Sec.~{\ref{sec:formalism}), explain the choice of parameters (Sec.~{\ref{sec:transmission}), analyze the properties of the transmission (Sec.~{\ref {sec:spin-density}), derive expressions and present results for spin density and spin current density (Sec.~{\ref{sec:scdo}), and conclude with a short summary (Sec.~{\ref{sec:summary}). Some technical steps and additional figures are relegated to the Appendix {\ref{sec:appendix}. 

\section{Formalism} \label{sec:formalism}

Consider massless 2D Dirac electrons in SLG (lying in the $x$-$y$ plane) that are scattered from a 1D rectangular potential barrier of $u(x)=u_0 \Theta(x)\Theta(d-x)$ and subject to a uniform electric field ${\bf E}=E_0 \hat{\bf z}$. Our treatment is carried out within the continuum formulation near one of the Dirac points, say ${\bf K}'$, and focuses on forward scattering, $k_y=0$,  (rendering the problem to be one dimensional). Recall that, in addition to the isospin ${\bm \tau}$ encoding the two-lattice structure of graphene, there is now a {\it real spin}, ${\bm \sigma}$. The pertinent Hamiltonian (in units where the kinetic energy parameter $\hbar v_F = 1$) is
\begin{eqnarray} \label{1}
&& h(-i \partial_x,\lambda) = [-i  \partial_x + \lambda (\hat {\bf z} \times {\bm \sigma})_x] \tau_x + \nonumber \\
&& \lambda (\hat {\bf z} \times {\bm \sigma})_y \tau_y +u(x)  \equiv h_0(-i \partial_x,\lambda)+u(x),   
\end{eqnarray}
which is a 4$\times$4 matrix first-order differential operator in ${\bm \sigma}\otimes{\bm \tau}$ space acting on the four component wave function $\psi(x)$ subject to scattering boundary conditions.  The wave function $\psi(x)$ is expressible as a combination of plane-wave vectors $e^{\pm i k_x x}v(\pm k_x)$  for $x \notin [0,d])$, and $e^{\pm i q_x x}w(\pm q_x)$ for $x \in [0,d]$. Here $k_x, q_x,\veps, u_0, \lambda$ are expressed in (nm)$^{-1}$ and $d$ is expressed in nm.  However, in presenting our numerical results below, energies $\veps, u_0$ and RSOC strength $\lambda$ will be presented in meV, [1 (nm)$^{-1} = 659.107$ meV].  The 4 component vectors $v(\pm k_x)$ and $w(\pm q_x)$ have unit current density $v^\dagger \hat{j_x} v= w^\dagger \hat{j_x} w=1$ where $\hat{j}_x=\frac{1}{A}I_{2 \times 2} \otimes \tau_x$ is the current operator and $A$ is an area unit. They satisfy the equations,
\begin{eqnarray} \label{3}
 && h_0(\pm k_x,\lambda)v(\pm k_x)=\veps v(\pm k_x), 
 \nonumber \\
&&
 h(\pm q_x,\lambda)w(\pm q_x)=\veps w(\pm q_x).
\end{eqnarray} 
Because RSOC acts in all of space (not only in the barrier region), the vectors $v(\pm k_x)$ cannot be chosen as spin eigenfunctions {\it since spin is not conserved}.  Equations~(\ref{3}) are not eigenvalue equations.  Rather, they are implicit equations for the wave numbers $k_x(\veps)$ and $q_x(\veps)$ which depend on the fixed scattering energy as well as for $u[\pm k_x(\veps)]$ and $w[\pm q_x(\veps)]$. For $\veps>0$, there are two wave numbers in each region that solve these implicit equations, $\pm k_{xn}(\veps)$ for $x \notin [0,d]$, and $\pm q_{xn}(\veps)$ for $x \in [0,d]$ ($n$=1,2).  Solution of Eqs.~(\ref{3}) yields (see SM)
\begin{eqnarray} \label{3kn}
 && k_{xn}^2=[\veps+(-1)^{n+1} \lambda]^2-\lambda^2, \nonumber \\
 && q_{xn}^2=[\veps+(-1)^{n+1} \lambda-u_0]^2-\lambda^2.
\end{eqnarray}
The wave function corresponding to an incoming wave in channel $n$ ($n=1,2$) in the three regions is,
\begin{equation} \label{4}
\psi_n(x)=\begin{cases} \underbrace{\vert k_{xn}\ra+r_{n1} \vert \bar{k}_{x1} \ra+r_{n2} \vert \bar{k}_{x2} \ra}_{x<0}\\ \underbrace{a_{n1}^+\vert q_{x1}\ra+a_{n2}^+\vert q_{x2}\ra +
a_{n1}^- \vert \bar{q}_{x1} \ra + a_{n2}^- \vert \bar{q}_{x2} \ra}_{0<x<d}\\ \underbrace{t_{n1} \vert k_{x1} \ra+t_{n2} \vert k_{x2} \ra}_{x>d}, \end{cases}
\end{equation}
where $\vert k_{xn}\ra \equiv e^{i k_{xn}x} v_n(k_{xn})$, $\vert \bar{k}_{xn}\ra \equiv \vert -k_{xn}\ra$, $\vert q_{xn}\ra \equiv e^{i q_{xn}x} w_n(q_{xn})$, and $\vert \bar{q}_{xn}\ra \equiv \vert -q_{xn} \ra$. 

The matching conditions at $x=0$ and $x=d$ yield the transmission and reflection amplitude matrices $t = \binom{t_{11} t_{12}}{t_{21}t_{22}}$ and $r=\binom{r_{11} r_{12}}{r_{21} r_{22}}$ together with the eight coefficients $\{a_{nm}^\pm \}, (n,m=1,2)$.  The wave functions $\psi_n(x)$ with scattering boundary conditions are therefore determined everywhere. The solution obeys unitarity and current conservation,
\begin{eqnarray}
&& T+R=\mbox{Tr}[t^\dagger t]+\mbox{Tr}[r^\dagger r]=2, \nonumber \\
&& \frac{d}{dx} j(x)\equiv \frac{d}{dx} \sum_{n=1}^2 \psi_n^\dagger(x) \hat{j}_x \psi_n(x)=0.
 \label{5}
\end{eqnarray}
$T$ and $R$ are the transmission and reflection coefficients and $\hat{j}_x=I_2 \otimes \tau_x$ is the current operator defined after Eq.~(\ref{3}). 

The numerical choice of parameters is dictated by experiments.  In Ref.~\cite{Gmitra}, it is shown that for a field $E_R=2$ V/nm, $\lambda$ is on the order of a fraction of an meV.  Here we let $0<\lambda \le 0.659$ meV, and vary the barrier height in the range $0<u_0<200$.  The barrier width $d$ is taken at $200$ nm (except for Fig.~\ref{Fig1} where $d=60$ nm). 
Concerning the choice of $\veps$, it is expected that the interesting physics occurs for $\veps$ close to $\lambda$. The reason is that the RSOC partially lifts the spin degeneracy, and at the Dirac point the energy splitting is of order $\lambda$. The role of electron spin is relevant when the scattering energy is close to the two split levels (see Fig.~8(b) in Ref.~\cite{Richter}). 

\section{Transmission} \label{sec:transmission}

For $k_y=0$, the two channels are uncoupled, and the transmission coefficient is obtained analytically:
\begin{eqnarray} \label{7}
&& T(u_0;\veps,\lambda)= \\
&& \sum_{n=1}^2 \frac{q_{xn}^2}{(u_0-\veps)[(u_0-\veps)+4 \lambda]-\lambda^2 \sin^2(q_{xn}d)} .\nonumber
\end{eqnarray}
A necessary condition for the occurrence of Klein tunneling is $k_n^2>0, n=1,2$$\ \Rightarrow \veps>2 \lambda>0$.  {\it Full Klein tunneling} occurs if $q_n^2>0, n=1,2$, i.e., $u_0>\veps+2 \lambda$. If $(q_1q_2)^2<0$, i.e., $\veps>u_0>2 \lambda >0$ (where $q_{x1}^2>0, q_{x2}^2<0$) or $\veps+2 \lambda>u_0>\veps-2\lambda>0$, (where $q_{x1}^2<0, q_{x2}^2>0$) we have {\it Partial Klein tunneling}.  If $q_{xn}^2<0$, the corresponding denominator in Eq.~(\ref{7})
is very large that the contribution of this term to the transmission is minuscule, because $-\sin^2 (q_{xn} d) = \sinh^2(\vert q_{xn} \vert d) \gg 1$ and channel n is virtually closed. Thus, Klein tunneling is still manifest even if only one channel is open, 
but then $T \le 1$. 
Inspecting $T(u_0;\veps,\lambda)$ as function of $u_0$ in Fig.~\ref{Fig1}, shows that  when the scattering energy equals the barrier height, ($\veps = u_0$ where channel 1 is about to close and channel 2 is about to open), $T(u_0;\veps=u_0,\lambda) \approx 0$. Recall from Eq.~(\ref{7})  that for $\lambda=0$ the transmission at the forward direction is unimpeded,  $T(u_0;\veps,\lambda=0)=2$ (identically). But the inset in Fig.~\ref{Fig1} shows that this happens for very small $\lambda$. Note that in the corresponding 1D Schr\"odinger problem for $\veps=u_0$,  $T=\frac{4}{4+(kd)^2}$ where $k=\sqrt{\frac{2 m \veps}{\hbar^2}}$.  Therefore, an experiment measurement of transmission for $\lambda \ne 0$ should be an excellent probe of the strength of RSOC in SLG.

\begin{figure}[htb]
\centering
{\includegraphics[width=0.9\linewidth]{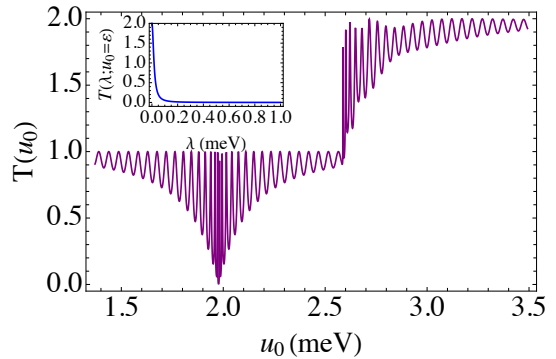}} 
\caption{\footnotesize 
Partial Klein tunneling ($T<1$) and full Klein tunneling ($1<T<2$) through a barrier of small height $u_0$, occurs when $\vert u_0-\veps \vert < 2 \lambda$ (partial) and $u_0>\veps+2\lambda$ (full).  For $\veps=1.977$  meV, $\lambda = 0.30314$ meV, and $d=60$ nm, 
$T$ is plotted {\it v.s} $u_0$ in the range $\veps-2 \lambda <u_0<3.5$ meV.   A remarkable feature of RSOC is that for $u_0 \approx \veps$, (where channel 1 is about to close and channel 2 is about to open), $T$ nearly vanishes expect for $\lambda \to 0$ where, as shown in the inset, it sharply rises to 2 [see Eq.~(\ref{7})].  
}
\label{Fig1}
\vspace{-0.2in}
\end{figure}

We now consider the transmission coefficient $T$ and the (matter) current $j(x)$ as a function of potential barrier height parameter $u_0>\veps+2 \lambda$ and the RSOC strength parameter $\lambda$.  Figure \ref{Fig2} shows the transmission and current versus $\lambda$ for fixed $u_0$ 
 and Fig.~1(a) in the SM shows the transmission and current versus $u_0$ for fixed $\lambda$.  The main conclusion from these figures is that in the presence of RSOC, the transmission coefficient is no longer unimpeded. Rather, for fixed $u_0$ and for experimentally relevant interval $0<\lambda<0.65$ meV the transmission smoothly decreases as in Fig.~\ref{Fig1}. And for fixed $\lambda$, considered as function of the barrier height $u_0$  it shows a pattern of oscillations below the unitarity upper limit $T=2$ as shown in Fig.~1(a) in the SM. It is of course not surprising that the current and the transmission coefficient are highly correlated. Note that the current is space-independent [see before Eq.~(\ref{3})]. 
  
\begin{figure}[htb]
\centering
{\includegraphics[width=0.9\linewidth]{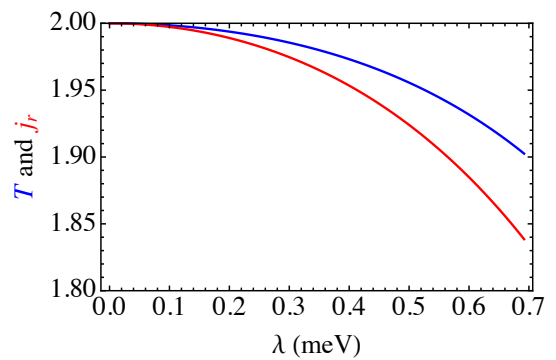}} 
\caption{\footnotesize 
Transmission $T$ (blue) and current $j_x$ (red) for the p-n-p junction, 
versus $\lambda$ for $u_0=195$ meV, $d=200$ nm, $\veps=6.65$ meV and $k_y=0$ (recall that for $\lambda=0$, $T(k_y=0)=2$ identically.   
}
\label{Fig2}
\end{figure}

\section{Spin density}  \label{sec:spin-density}

Spin density (and spin current density) operators $\{ {\cal O} \}$ are representable as 4$\times$4 matrices in ${\bm \sigma} \otimes {\bm \tau}$ space. Spin observables are obtained as $O(x)=\psi^\dagger(x) {\cal O}\psi(x)$ (this is not an expectation value, spin-observables may depend on $x$). The spin density operators $\hat{\bf S}$ and the spin density observables ${\bf S}(x)$ are given by, 
\begin{eqnarray} \label{Sop}
&& \hat{\bf S}
=\tfrac{1}{2} \hbar {\bm \sigma} \otimes {\bf I}_2,
\ \ {\bf S}(x)=\frac{1}{4} \sum_{n=1}^2 \psi^\dagger_n(x) \hat{\bf S} \psi_n(x),
\end{eqnarray} 
where $\psi_n(x)$ is defined in Eq.~(\ref{4}). The unit of spin density used here is $S_0=\hbar/A$.
As expected, $S_x=S_z=0$, because the polarization is proportional to ${\bf k} \times {\bf E} \parallel \hat{\bf y}$, and here  $k_y=0$.
Figure~\ref{Fig3} shows the spin density $S_y$ as a function of $\lambda$ (spin densities are space-independent). The polarization increases rapidly for $\lambda >  0.20$ meV. Figure~1(b) in the SM plots $S_y$ versus $u_0$, and shows a rich oscillatory pattern that decreases near the lower limit $u_0 =\veps + 2 \lambda$, below which one channel is closed (see discussion of Fig.~\ref{Fig1}). Both figures substantiate the role of $\lambda$ and $u_0$ as useful parameters to control the degree of polarization.

\begin{figure}[htb]
\centering
{\includegraphics[width=0.9\linewidth]{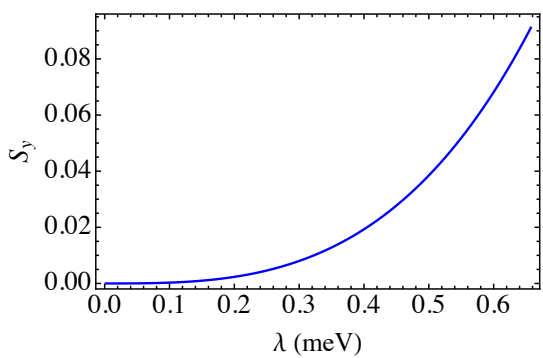}} 
\caption{\footnotesize Spin density $S_y$ as function of $\lambda$ for $u_0$=192 meV, $d=200$ nm, $\veps=2.60$ meV and $k_y=0$.
}
\label{Fig3}
\vspace{-0.1in}
\end{figure}

\section{Spin current density operators and observables}  \label{sec:scdo}

The (tensor) spin-current density operator $\mathbb{J}$ and the observed components of the spin current density $J_{ij}(x)$ are defined as
\begin{eqnarray} \label{Jop}
&& \hat{\mathbb{J}}=\frac{1}{2} \{ \hat{\bf S},\hat{\bf V} \}, 
\Rightarrow \ \hat{\mathbb{J}}_{i,j}=\hat{S}_i\hat{V}_j+\hat{V}_j\hat{S}_i, \nonumber \\
&& J_{ij}(x)=\frac{1}{2} \sum_{n=1}^2 
\psi^\dagger_n(x) \hat{\mathbb{J}}_{ij} \psi_n(x), 
\end{eqnarray} 
where $\hat{\bf S}$ is the spin density operator defined in Eq.~(\ref{Sop}) and $\hat{\bf V} = {\bf I}_2 \otimes {\bm \tau}$ is the velocity operator [$= \hat{j}_x$ defined before Eq.~(\ref{3})].  In Eq.~(\ref{Jop}), $i=1,2,3 = x,y,z$ specifies the polarization direction, and $j=1,2=x,y$ the direction of propagation.  The unit of spin current density is $J_0=S_0 v_F=\gamma /A = 659.107$ meV/nm.

\begin{figure}[htb]
{\includegraphics[width=0.9\linewidth]{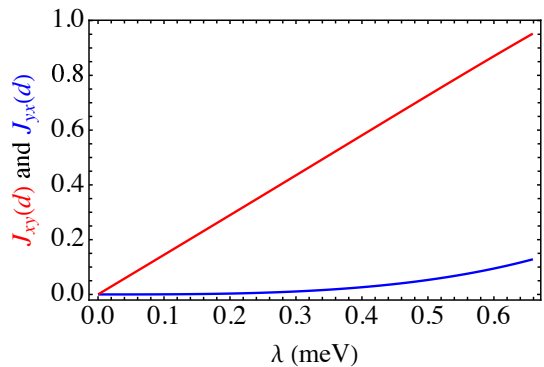}}
\caption{\footnotesize 
Spin current densities $J_{yx}$ and $J_{xy}$   (in units of $J_0$) 
versus $\lambda$ 
at the right wall of the barrier ($x=d$), versus $\lambda$ for $u_0$=192 meV, $d=200$ nm, $\veps$=2.60 meV. Note that the symmetry condition $J_{xy}=-J_{yx}$ valid in bulk SLG \cite{Zhang} is broken by the 1D potential barrier.  
}
\label{Fig4}
\vspace{-0.1in}
\end{figure}

The spin current density was calculated in bulk SLG in Ref.~\cite{Zhang}; it was found that 
(1) $J_{xx}=J_{yy}=J_{zx}=J_{zy}=0$, 
(2) $J_{xy}=-J_{yx}$, and
(3) The spin currents are independent on space [see Eq.~(5) in Ref.~\cite {Zhang}].
Below, we show that: 
(1) In the presence of a 1D potential (where there is no rotational symmetry around the $z$-axis), the symmetry relation $J_{xy}=-J_{yx}$ that is valid in bulk SLG \cite{Zhang} is broken. 
(2) Although the value of $\lambda$ used in our calculations is much smaller than that used in Ref.~\cite{Zhang}, the size of spin current densities
 are both systems is the same.  
(3) The spin current densities are not uniform, and the continuity equation 
includes a spin torque density term \cite{Niu} (see below). 
Figure~\ref{Fig4} shows $J_{xy}(d)$ and $J_{yx}(d)$ as function of $\lambda$. Note the nearly linear increase of {\it e.g} $J_{xy}(d)$ with $\lambda$. No symmetry exists between $J_{xy}$ and $J_{yx}$. 
Figure~(2a) in the SM shows the spin current densities $J_{xy}(0)$ (red curve) and $J_{yx}(0)$ (blue curve), just at the left wall of the barrier, for the p-n-p junction as function of $\lambda$. The size of the spin current density $J_{xy}(0)$ (electrons polarized along $x$ and propagate along $y$) for $\lambda \approx$ 0.659 meV is indeed large (as compared with $J_{xy}$ in bulk SLG  calculated for $\lambda \approx$ 45 meV).  

To stress the role of the Klein paradox in the present system we compare our results with those obtained in bulk SLG (where there is no Klein paradox). In Eq.~(7) of Ref.~\cite{Zhang}, the authors found that in bulk SLG, $J_{yx}=J_0 \frac{\eta^2+\cos 2 \phi}{1+\eta^2}$ where $\eta=\frac{\veps}{\sqrt{k_x^2+k_y^2}}$ is of order unity and $\phi=\arctan k_y/k_x$. Thus, for $k_y =0$, this implies $J_{yx}=-J_{xy} \approx J_0$. As shown in Figs.~\ref{Fig4}, \ref{Fig5} and \ref{Fig6}, in the presence of 1D potential barrier, the spin current density can reach similar values. However, in Ref.~\cite{Zhang} the value of $\lambda$ is about 200 times higher than the one we have used.  As shown in Ref.~\cite{Gmitra}, such high values of $\lambda$ are not achievable in SLG. It is possible to have higher values of RSOC if the SLG is in contact with metals such as Au or Pb. But upon passing a current through the metallic substrate, the electrons will flow through the metal and not through the SLG, so at the Fermi level, the electronic states are metallic. 

Spin current densities versus $u_0$ are shown in Fig.~\ref{Fig5}. Recalling the behavior of $S_y(u_0)$  shown in Fig.~1(b) in the SM, and of $J_{xy}(u_0)$ and $J_{yx}(u_0)$ in Fig.~\ref{Fig5}, it is clear that the spin density and the spin current density are significantly correlated. 

\begin{figure}[htb]
\centering
{\includegraphics[width=0.9 \linewidth]{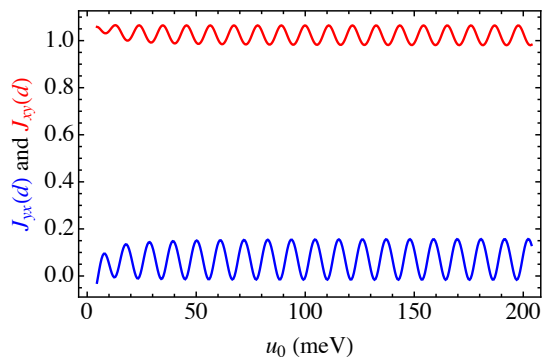}} 
\caption{\footnotesize  Spin current densities $J_{xy}(d), J_{yx}(d)$ on the right wall of the barrier versus $\lambda$ for $\veps+2 \lambda<u_0<200$ meV, $d=200$ nm, $\veps$=2.60 meV, $\lambda$=0.695 meV and $k_y=0$.  
}
\label{Fig5}
\end{figure}

\begin{figure}[htb]
\centering
{\includegraphics[width=0.9\linewidth]{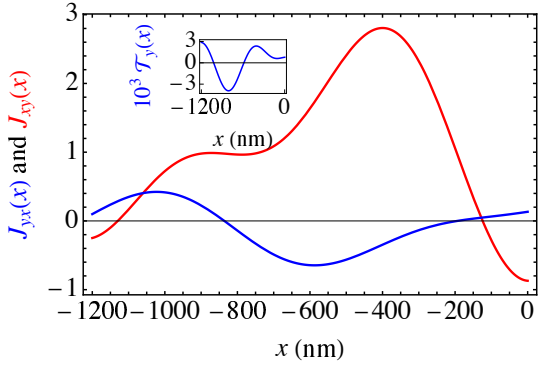}} 
\caption{\footnotesize Space dependence of the spin current densities $J_{xy}(x)$ and $J_{yx}(x)$ (in units of $J_0$) for $-1200<x<0$ nm (region to the left of the barrier, where the wave is reflected) for $\lambda=0.695$ meV, $\veps=2.60$ meV, and $u_0=198.2$ meV. Oscillations due to the linear combination of plane waves continues to $x \to -\infty$.  The inset shows the spin torque density ${\cal T}_y(x)$ in units of $J_0$/nm.  }
\label{Fig6}
\end{figure}

Consider now the space dependence of the spin current densities $J_{xy}(x)$ and $J_{yx}(x)$  (recall that the spin density ${\bf S}$ is space-independent).  They are plotted in Fig.~\ref{Fig6} to the left of the barrier (the reflected region) where the wave numbers $k_{xn}$ [defined in the first row of Eq.~(\ref{3kn})], are small (since $\veps$ and $\lambda$ are small).  This space dependence results in non-zero divergence, and the corresponding continuity 
equation requires the inclusion of finite spin torque density ${\cal T}_i, (i=x,y,z)$ \cite{Niu}.
Following the discussion in section III of the SM, there are two relevant vector fields ${\bf J}_1(x) = (0,J_{xy}(x))$ and ${\bf J}_2(x) = (J_{yx}(x),0)$ that satisfy continuity equations ${\bm \nabla} \cdot {\bf J}_i = {\cal T}_i$. Therefore, in our case, the only non-zero spin torque density is ${\cal T}_2(x) = dJ_{yx}(x)/dx$. As shown in Fig.~\ref{Fig6}, in the region $x<0$, $J_{yx}(x)$ is smooth, and the spin torque density derived from it is well defined (and measurable).

\section{Summary and Conclusion}  \label{sec:summary}

The Klein  paradox in SLG occurs at the Fermi energy $\veps$ when an electron tunnels through a 1D potential barrier of height $u_0$ (which can be experimentally controlled by a gate voltage) in the region $u_0>\veps>0$. When, in addition, a uniform perpendicular electric field ${\bf E} = E_0 \hat {\bf z}$ is applied, the role of electron spin enters due to RSOC. Here we elucidate the physics when the Klein paradox and RSOC are combined, in order to expose interesting facets of graphene spintronics within a time-reversal invariant formalism. The fact that $u_0$ and $\lambda$ can be easily experimentally controlled makes our analysis readily verifiable.
 
Our main results can be summarized in terms of the figures.
(1) Transmission coefficient $T$: Due to partial Klein tunneling, $0<T<1$ for $\vert u_0-\veps \vert<2 \lambda$ and $T \approx 0$ for $\veps = u_0$ and $\lambda>0.2$ meV (Fig.~\ref{Fig1}). For $u_0 > \veps+2 \lambda$ The transmission decreases smoothly as function $\lambda$ (Fig.~\ref{Fig2}). Recall that $T=2$ for $\lambda=0$ and $k_y=0$.
(2) Spin density: $S_x=S_z=0$ and, as shown in Fig.~\ref{Fig3}, the space independent spin density $S_y(\lambda)$ increases rapidly with $\lambda$.
(3) Spin current densities $J_{ij}(x)$: As shown in Figs.~\ref{Fig4}, \ref{Fig5} and \ref{Fig6}:
the symmetry $J_{xy}=-J_{yx}$ in bulk SLG is broken, $J_{xy}$ increases linearly with $\lambda$, and both $J_{xy}(x)$ and $J_{yx}(x)$ are space dependent and the continuity equation for $J_{yx}(x)$ includes a non-zero spin torque density \cite{Niu}. 
Clearly, in the present system (as compared with bulk SLG), spin density and spin current density have a much richer behavior.

Complimenting the developments in spintronics \cite{David, Spintronics}, our work is partially motivated by the quest for constructing spintronic devices without the use of an external magnetic field or magnetic materials  \cite{Hatano, Matityahu, AB1}. Specifically for graphene, it shows the vanishing of the transmission at energy $\veps=u_0$, the linear dependence of spin current density $J_{xy}(d)$ on $\lambda$, and the occurrence of spin torque density, thereby advancing this goal.

\begin{acknowledgments}
Discussions with J. Nitta, J. Fabian, K. Richter, M. H. Liu and S. Ilani were indispensable for understanding some crucial issues. 
\end{acknowledgments}

\appendix

\section{}  \label{sec:appendix} 

Here we briefly expand on several points discussed in the main text. Recall that the pertinent system consists of electron scattering in single layer graphene at Fermi energy $\veps$ that undergoes Klein tunneling through a 1D rectangular barrier of height $u_0$, width $\ell$ and subject to a Rashba spin-orbit coupling (RSOC) of strength $\lambda$.  The main topics clarified below include: (1) Derivation of Eqs.~(\ref{3}) and (\ref{3kn}).  (2)  Additional details on the occurrence of {\it Partial Klein tunneling} in the region $\veps-2 \lambda \le u_0 \le \veps + 2 \lambda$ and its dramatic effect on the transmission in this region.  (3) Additional figures showing the spin current density as a function of the barrier height $u_0$ and the RSOC strength $\lambda$. (4) Discussion of the space dependence of spin current density, the pertinent continuity relation and its relation to the spin torque density \cite{Niu}.

\subsubsection{Equations (\ref{3}) and (\ref{3kn})} \label{SMA}

The precise form of $h_0(k_x,k_y,\lambda)$ in Eq.(\ref{3}), in the general case $k_y \ne 0$, is
\begin{equation} \label{SM1}
h_0(k_x,k_y,\lambda)=\begin{pmatrix} 
0&k_x-ik_y&0&0\\
k_x+ik_y&0&2i\lambda&0\\
0&-2i \lambda&0&k_x-ik_y\\
0&0&k_x+ik_y&0 
 \end{pmatrix} .
\end{equation} 
Its two positive eigenvalues are
\begin{equation} \label{SM2}
\veps_{1,2}=\mu(k_x,k_y,\lambda) \mp \lambda,
\end{equation} 
where
\begin{equation}
\mu(k_x,k_y,\lambda) \equiv \sqrt{k_x^2+k_y^2+\lambda^2}.
\end{equation}
The corresponding eigenvectors are, 
\begin{equation} \label{SM3}
v_{1,2}(k_x,k_y)=\frac{1}{2}\sqrt{1\pm \frac{\lambda}{\mu(k_x,k_y,\lambda)}}
\begin{pmatrix} \mp ik_x+k_y\\i[\lambda \mp \mu(k_x,k_y,\lambda)]\\
\mp \lambda+\mu(k_x,k_y,\lambda) \\ k_x+ik_y \end{pmatrix}. 
\end{equation}
For $k_y=0$ (forward propagation) and fixed scattering energy $\veps$, this means that the energies in Eq.~(\ref{SM2}) are both equal, $\veps_1=\veps_2=\veps$. Solving for $k_x$ in each equation yields the two wave numbers $\{k_{xn}(\veps)\}$ ($n=1,2$) specified in the first equation \ref{3kn}. These wave numbers, when inserted in the vectors $v_{1,2}(k_x)$ defined in Eq.~(\ref{SM3}) above determine the spinors $\{v(k_{xn})\}$ used in Eq.~(\ref{4}). A similar procedure applies for the wave vectors $\{q_{xn}(\veps)\}$ and the spinors $w_n(q_{xn})$.  They are obtained respectively from $k_{xn}$ and $\{v(k_{xn})\}$ after replacing $\veps \to \veps-u_0$. 

\subsubsection{Transmission through a low barrier  and partial Klein tunneling}
\label{SMB}

There are various configurations related to Klein transmission through a barrier of height $u_0$ in the presence of Rashba spin-orbit coupling of strength $\lambda$, determined by inequalities involving $u_0, \veps$ and $\lambda$. (1) For $0<u_0<\veps-2\lambda < \veps$  [to the left of the blue point in Fig.~\ref{SMFig1}(a) and $1<T<2$ in the left of Fig.~\ref{SMFig1}(c)], we have scattering above the barrier with two channels open $q_{x1}^2, q_{x2}^2>0$ and there is no tunneling. (2) For $\veps-2\lambda <u_0< \veps$   [between the blue and red points in Fig.~\ref{SMFig1}(a), panel 2 in Fig.~\ref{SMFig1}(b) and $0<T<1$ in the left of Fig.~\ref{SMFig1}(c)] ], $q_{x1}^2>0, q_{x2}^2<0$ we have scattering above the barrier with one channel open. This is partial Klein tunneling above the barrier. (3) For $\veps <u_0< \veps+2\lambda $   [between the red and green points in Fig.~\ref{SMFig1}(a), panel 4 in Fig.~\ref{SMFig1}(b), and $0<T<1$ in the right of Fig.~\ref{SMFig1}(c)] $q_{x1}^2<0, q_{x2}^2>0$ we have scattering below the barrier with one channel open. This is partial Klein tunneling below the barrier. (4) For $0<\veps+2\lambda<u_0$ [to the right of the green point in Fig.~\ref{SMFig1}(a) and panel 5 in Fig.~\ref{SMFig1}(b) and $1<T<2$ in the right of Fig.~\ref{SMFig1}(c)] $q_{x1}^2, q_{x2}^2>0$ and we have scattering below the barrier with two channels open. This is full Klein tunneling below the barrier.

\begin{figure}[htb]
\centering
\subfigure[]
{\includegraphics[width=0.75\linewidth]{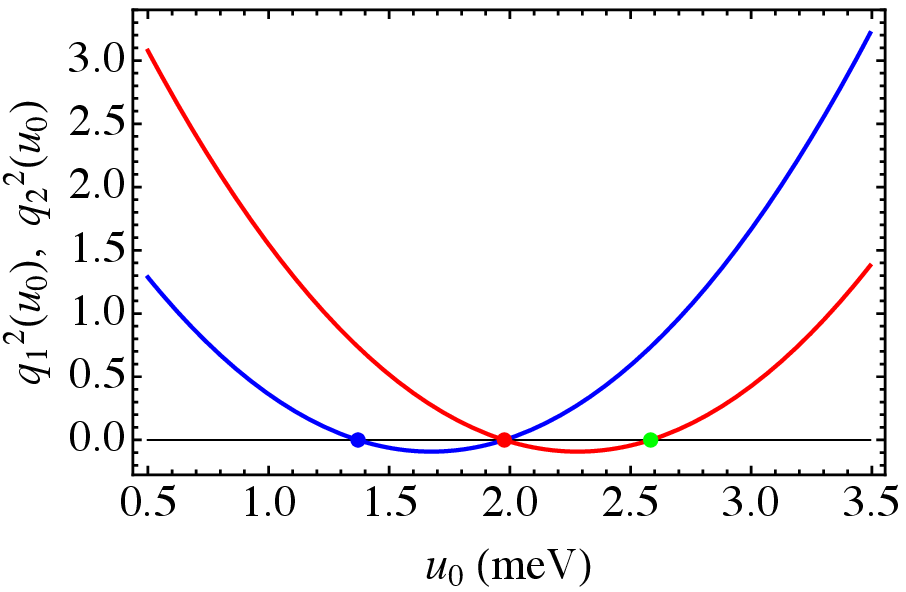}}
\centering
 \subfigure[]
{\includegraphics[width=0.85\linewidth]{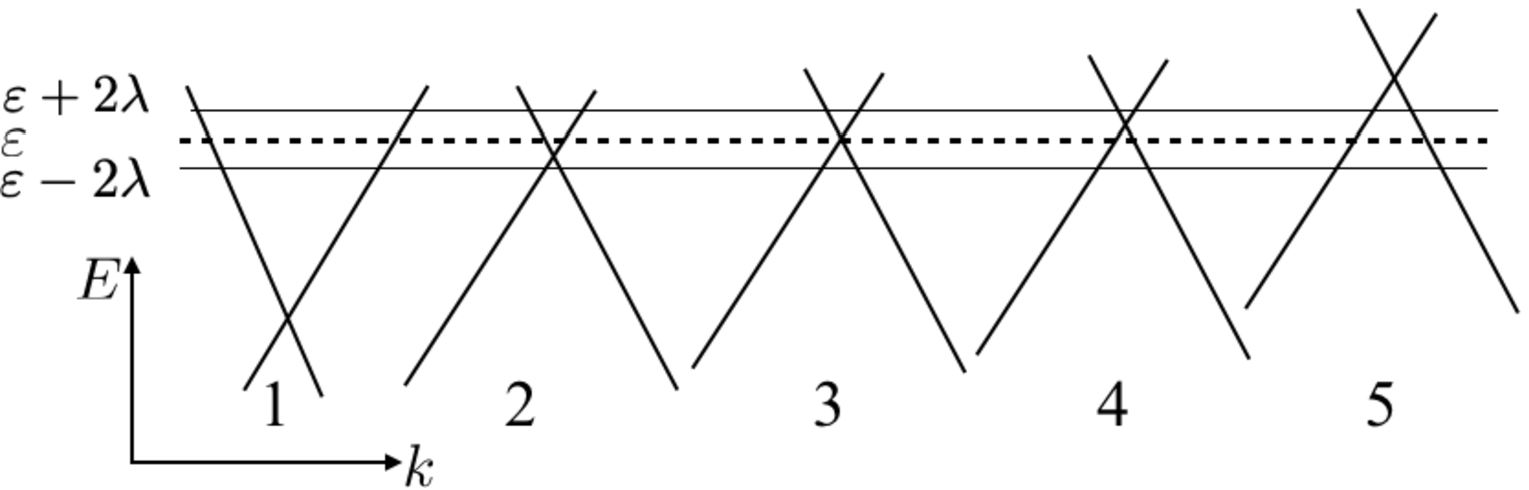}} 
 \subfigure[]
{\includegraphics[width=0.75\linewidth]{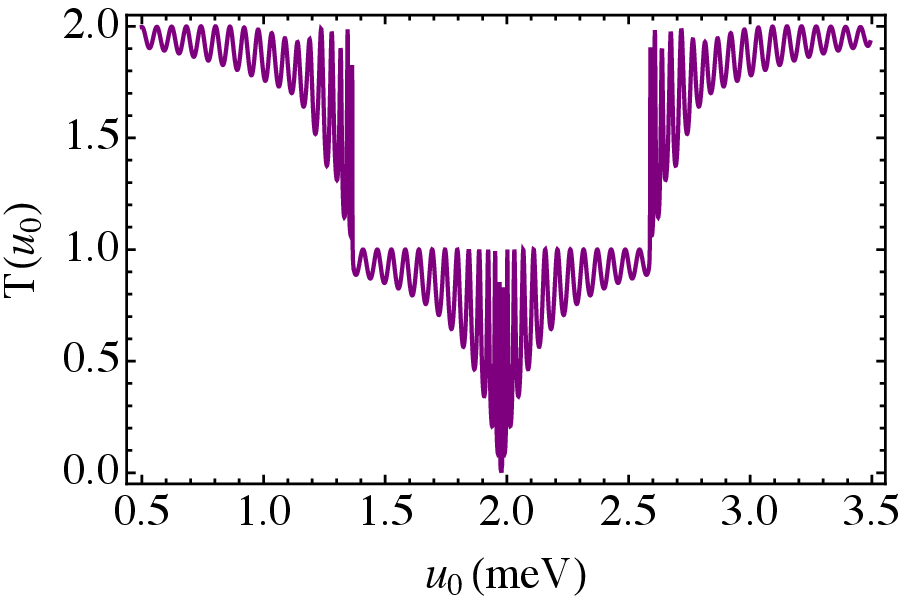}} 
\caption{\footnotesize 
Various configurations related to Klein transmission through a barrier of height $u_0$, in the presence of Rashba spin-orbit coupling of strength $\lambda$.  when $\vert u_0-\veps \vert < 2 \lambda$.  Here we take $\veps=1.977$  meV, $\lambda = 0.30314$ meV, and barrier width $d=60$ nm. (a) Squared wavenumbers, $q_{xn}^2$ defined in Eq.~(\ref{3kn}) (blue for $n=1$ and red for $n=2$). The blue, green and red points mark $u_0=(\veps-2 \lambda , \veps, \veps+2 \lambda)$. Negative $q_{xn}^2$ implies that channel $n$ is closed (does not contribute to the transmission). (b) Dirac cones for various values of $u_0$ (schematic), Fermi energy $\veps$ and SO splitting $\veps \pm 2 \lambda$ (magnified): In panels 1-5, $u_0=0, \veps-\lambda, \veps, \veps+ \lambda$ and $u_0>\veps+2 \lambda$.  (c) The total transmission is plotted versus $u_0$ for $\veps-2 \lambda <u_0<3.5$ (meV).  A remarkable feature of RSOC is that for $u_0 \approx \veps$ (red point in (a) where one channel is about to open and the other is about to close), the transmission nearly vanishes expect very near $\lambda$, as shown in the inset of Fig.~\ref{Fig1}, where it sharply rises to 2 [see Eq.~(\ref{Sop})].  
}
\label{SMFig1}
\end{figure}

\subsubsection{Spin current densities as function of $\lambda$ and $u_0$}
\label{SMC}
Here we display additional figures showing current densities as function of barrier height $u_0$ and RSOC strength $\lambda$. 
Fig.~\ref{SMFig2}(a) shows the transmission and the charge current for fixed $\lambda$ versus $u_0$ in the region $u_0>\veps+2 \lambda$ where there is full Klein tunneling. From Eq.~(\ref{7}) in it is evident that the pattern of the transmission is that of small oscillations (determined by the trigonometric term in the denominator) just below the maximal value $T=2$. It is  also reasonable to expect that the current is correlated with the transmission (note however that the current is space independent while the transmission is defined only to the right of the barrier). Figure \ref{SMFig2}(a) and Fig.~\ref{SMFig1}(c) together with Fig.~\ref{Fig2} completes our visage of the transmission coefficient as function of $u_0$ and $\lambda$.  Figure~\ref{SMFig2}(b) plots the only non-zero spin density $S_y$ for fixed $\lambda$ as function of $u_0$ in the region $u_0>\veps+2 \lambda$ where there is full Klein tunneling.  Recall that $S_y$ is space independent. 
\begin{figure}
\centering
\subfigure[]
{\includegraphics[width=0.75\linewidth]{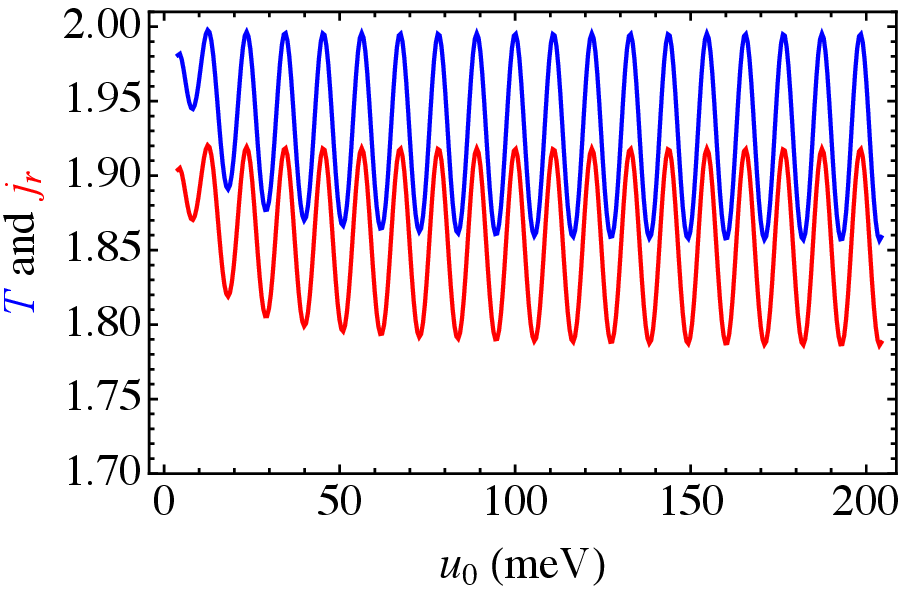}} 
\ \ \ \ \subfigure[]
{\includegraphics[width=0.75\linewidth]{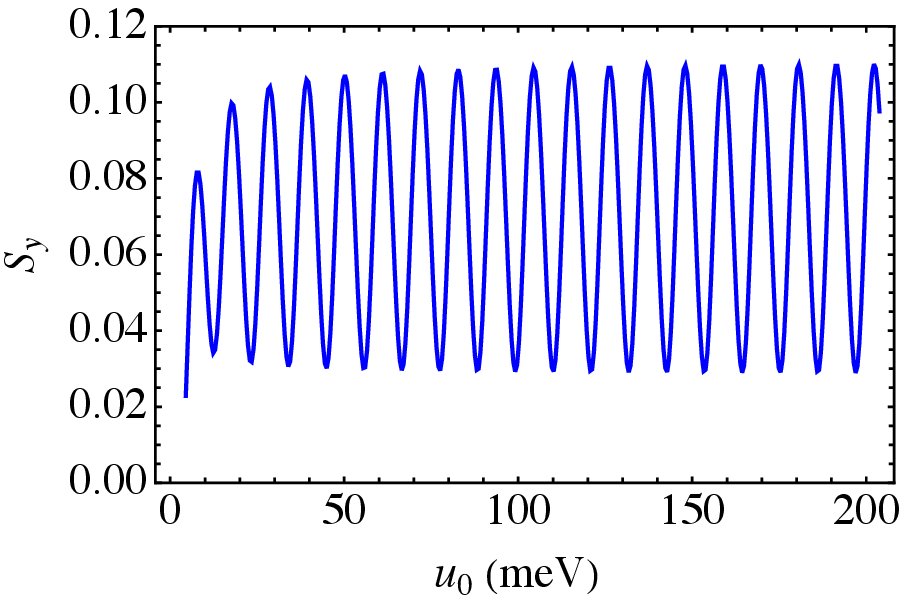}}  
\caption{\footnotesize (a) Transmission $T$ (blue) and current $j_x$ (red) for the p-n-p junction 
versus $u_0$ in the range $\veps+2 \lambda<u_0<200$ meV for $\lambda=0.695$ meV. Other parameters as in Fig.~\ref{Fig1}.  (b) Spin density $S_y$ versus $\veps+2 \lambda<u_0<200$ meV for $\lambda=0.695$ meV and $k_y=0$. Other parameters are as in Fig~\ref{Fig3}. Note that $S_y(x)$ is independent of $x$ throughout the sample.
}
\label{SMFig2}
\end{figure}

Figure~\ref{SMFig3}(a) shows the spin current densities $J_{xy}(0)$ (red curve) and $J_{yx}(0)$ (blue curve) versus $\lambda$, just to the left wall of the barrier for the p-n-p junction. In the main text, these quantities are are shown at $x=d$ (that is the right wall of the barrier). Both $J_{xy}(0)$ and $J_{xy}(d)$ are nearly linear with $\lambda$ but have different sign.  The size of the spin current density $J_{xy}(0)$ (electrons polarized along $x$ and propagate along $y$) for $\lambda \approx$ 0.659 meV is indeed large (same order of magnitude as in $J_{xy}$ calculated in bulk SLG for $\lambda \approx$ 45 meV, that is about two orders of magnitude higher than the value of $\lambda$ used here).  Figure \ref{SMFig3}(b) shows the spin current densities $J_{xy}(0)$ (red curve) and $J_{yx}(0)$ (blue curve) as function of $\lambda$ on the left wall of the barrier.  Compare with Fig.~\ref{Fig5}, where these quantities are shown at the right wall of the barrier at $x=d$.

\begin{figure}[htb]
\subfigure[]
{\includegraphics[width=0.75\linewidth]{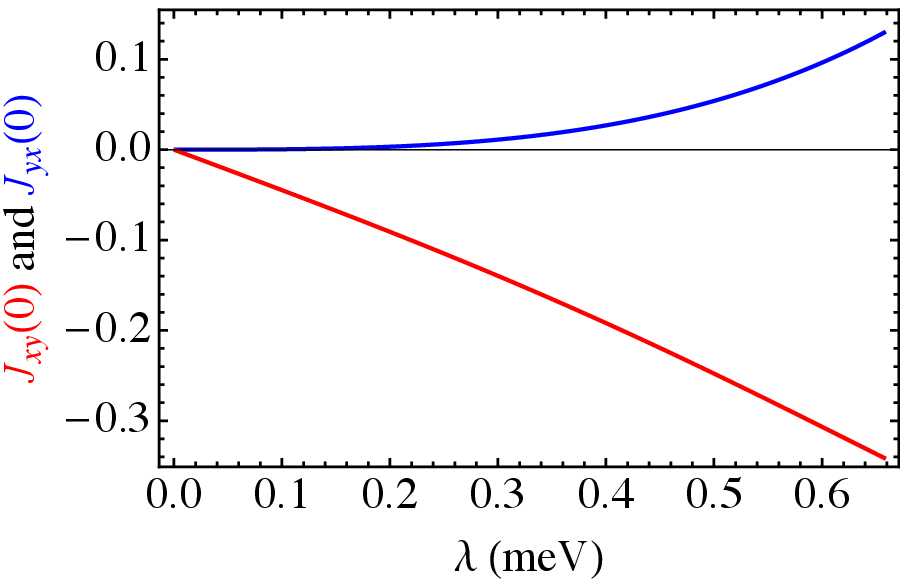}}
\ \ \ \ \ \subfigure[]
{\includegraphics[width=0.75\linewidth]{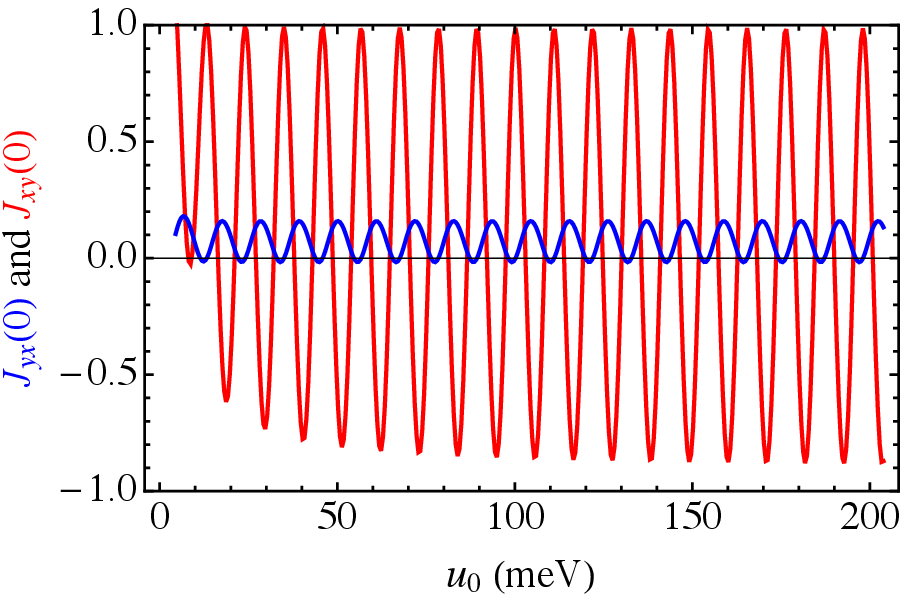}}
\caption{\footnotesize 
(a) Spin current densities $J_{yx}(0)$ and $J_{xy}(0)$   (in units of $J_0$) versus $\lambda$ at the left wall of the barrier ($x=0$), versus $\lambda$ for $u_0$=192 meV, $d=200$ nm, $\veps$=2.60 meV. Note that the symmetry condition $J_{xy}=-J_{yx}$ valid in bulk SLG is broken by the 1D potential barrier.  (b) Spin current densities $J_{xy}(0), J_{yx}(0)$ on the left wall of the barrier versus $\lambda$ for $\veps+2 \lambda<u_0<200$ meV, $d=200$ nm, $\veps$=2.60 meV, $\lambda$=0.695 meV and $k_y=0$. 
}
\label{SMFig3}
\end{figure}

\subsubsection{Space dependence of spin current density, continuity equation and spin torque}  \label{SME}

In the general case of a 3D material, the observable spin current density tensor ${\bf J}_{ij}(\bfr,t)$ depends on space and time. For each polarization direction $i=1,2,3=x,y,z$ the vector field,
\begin{equation} \label{SM4}
{\bf J}_i(\bfr,t) = [J_{ix}(\bfr,t), J_{iy}(\bfr,t), J_{iz}(\bfr,t)],
\end{equation} 
satisfies the continuity equation,
$$\frac{\partial S_i(\bfr,t)}{\partial t}+ {\bm \nabla} \cdot {\bf J}_i(\bfr,t)= {\cal T}_{i}({\bf r},t) \equiv \mbox{Re}[\psi^\dagger({\bf r},t) \hat{\cal T}_i \psi({\bf r},t)].$$
Here the scalar ${\cal T}_{i}({\bf r},t)$ is the spin torque density and $\hat{\cal T}_i=\frac{1}{i \hbar}[\hat{S}_i,H]$ is the spin torque density operator, where $H=({\bf p}+\lambda {\bm \sigma}\times {\bf n}_E)\cdot {\bm \tau}$ is the Hamiltonian operator and ${\bf n}_E$ is the unit vector in the direction of the electric field. The volume integral of ${\cal T}_{i}({\bf r},)$ sometimes vanishes due to symmetry relations \cite{Niu}, and then there exists a vector field ${\bf P}_i({\bf r},t)$ such that ${\cal T}_{i}({\bf r},)=-{\bm \nabla} \cdot {\bf P}_i({\bf r},)$. If the space dependence is just on $x$ the continuity equation is $\frac{\partial}{\partial t}S_i+ [\frac{d J_i(x)}{d x}+P_i(x)]=0$ and in the static case, this becomes 
\begin{equation} \label{SM5}
{\cal T}_{i}(x)=\frac {d J_i(x)}{d x}.
\end{equation} 
For 2D systems, with $J_{iz}(x)=0$, the nonzero  two component vector fields are
\begin{equation} \label{SM6}
{\bf J}_1(x)=[0,J_{xy}(x)], \ \ {\bf J}_2(x)=[J_{yx}(x),0].
\end{equation} 
 According to our discussion above, the only non-zero torque density scalar field is
\begin{equation} \label{SM7}
{\cal T}_{y}(x)=\frac {d J_{yx}(x)}{d x}.
\end{equation}

\onecolumngrid

\begin{thebibliography}{99}

\bibitem{Geim} 
K. S. Novoselov, A. K. Geim, S. V. Morozov, D. Jiang, Y. Zhang, S. V. Dubonos, I. V. Grigorieva, and A. A. Firsov, 
Science {\bf 306}, 666 (2004); A. K. Geim and K. S. Novoselov
Nature Mater. {\bf 6} 183 (2007); 
A. K. Geim
Science {\bf 324}, 1530 (2009). 

\bibitem{Guinea}
A. H. Castro Neto, F. Guinea, N. M. R. Peres, K. S. Novoselov, and A. K. Geim, 
Rev. Mod. Phys. {\bf 81}, 109 (2009).

\bibitem{Sarma}
S. Das Sarma, Shaffique  Adam, E. H. Hwang, and Enrico Rossi, 
Rev. Mod. Phys. {\bf 83}, 407 (2011).

\bibitem{Klein}
O. Klein,
Z. Phys. {\bf 53} 157 (1929);
C. W. J. Beenakker,
Rev. Mod. Phys. {\bf 80}, 1337 (2008). 

\bibitem{Katsnelson_06}
M. I. Katsnelson, K. S. Novoselov, and A. K. Geim,
Nature Physics {\bf 2}, 620 (2006). 


\bibitem{Katsnelson-2012}%
T. Tudorovskiy, K J. A. Reijnders and M. I. Katsnelson,
Phys. Scr. {\bf T146}, 014010 (2012).

\bibitem{Cheianov} Vadim V. Cheianov and Vladimir I. Fal’ko, 
Phys. Rev. {\bf B 74}, 041403 (2006). 


\bibitem{AF_11}
Pierre E. Allain and J-N. Fuchs, 
Eur.~Phys.~J.~B {\bf 83}, 301 (2011).

\bibitem{Ando}
T. Ando, and T. Nakanishi, 
J. Phys. Soc. Jpn. {\bf 67}, 1704  (1998).  

\bibitem{Peeters}
J. M. Pereira Jr., V. Molnar, F. M. Peeters, P. Vasilopoulos,
Phys. Rev. B {\bf 74}, 045424 (2006).

\bibitem{Barbier}
M. Barbier, P.  Vasilopoulos, and F. M. Peeters, 
Philosophical Transactions: Mathematical, Physical and Engineering Sciences
{\bf 368}, No. 193, 5499 (2010),  arXiv:1101.4117.

\bibitem{AB} 
Y. Avishai and Y. B. Band, 
Phys. Rev. {\bf B}102, 085435 (2020).
\bibitem{Huertas} H. D.Huertas, F. Guinea and A. Brataas,  
Phys. Rev. {\bf B 74}, 155426 (2006).

\bibitem{Min} H. Min, J.E. Hill, N.A. Shinitsyn, B.R. Sahu, L. Kleinman, and A.H. MacDonald, 
Phys. Rev. {\bf B 74}, 165310 (2006).
\bibitem{Yao}
Y. Yao, F. Ye, X. L. Qi, S. C. Zhang, and Z. Fang, 
Phys. Rev. {\bf B 75}, 041401(R) (2007).


\bibitem{Castro} E. V. Castro,  K. S. Novoselov, S. V. Morozov,  N. M. R. Peres,  J. M. B. L. dos Santos,  J. Nilsson, F. Guinea,  A. K. Geim,  and  A. H. Castro Neto, 
Phys. Rev. Lett. {\bf 99}, 216802 (2007). 

\bibitem{Trau}
B. Trauzettel, D. V. Bulaev, D. Loss, and G. Burkard, 
 Nature Phys. {\bf 3}, 192 (2007).
\bibitem{Tombros}
N. Tombros, C. Jozsa, M. Popinciuc, H. T. Jonkman, and B. J. van Wees, 
Nature {\bf 448}, 571 (2007).

\bibitem{Tombros1} N. Tombros, S. Tanabe, A. Veligura, C. Jozsa, M. Popinciuc, H. T. Jonkman, and B. J. van Wees,
 Phys. Rev. Lett. 101, 046601 (2008).
%

\bibitem{Cho} S. Cho, Y. Chen, and M. S. Fuhrer, 
 Appl. Phys. Lett. {\bf 91}, 123105 (2007).
\bibitem{Zarea}
M. Zarea and N. Sandler, 
Phys. Rev. {\bf B 79}, 165442 (2009).


\bibitem{Rashba1} 
E. I. Rashba,
Phys. Rev. {\bf B 79}, 161409(R) (2009).


\bibitem{Liu} M.-H. Liu and C.-R. Chang, 
Phys. Rev. {\bf B 80}, 241304(R) (2009). 

\bibitem{Gmitra0} M. Gmitra, S. Konschuh, C. Ertler, C. Ambrosch-Draxl, and
J. Fabian, 
Phys. Rev. B 80, 235431 (2009).

\bibitem{Berc}
D. Bercioux and De Martino, 
Phys. Rev. {\bf B 81}, 165410 (2010). 

\bibitem{Sergej1}
S. Konschuh, M. Gmitra, and J. Fabian,
Phys. Rev. {\bf B 82}, 245412 (2010).

\bibitem{Schwierz} F. Schwierz, 
 Nature Nanotechnol. {\bf 5}, 487 (2010).

\bibitem{Tse}
W.-K. Tse, Z. Qiao, Y. Yao  A. H. MacDonald, and Q. Niu, 
Phys. Rev. {\bf B 83}, 155447 (2011).

\bibitem{Sergej2} S. Konschuh, 
{it Spin-Orbit Coupling Effects From Graphene To Graphite}, Ph.D. Thesis, Universit\"at Regensburg, (2011).

\bibitem{Miao} G. Miao, M. Münzenberg, and J. S. Moodera, 
 Rep. Prog. Phys. {\bf 74}, 036501 (2011).
 
 \bibitem{Jo} S. Jo, D. Ki, D. Jeong, H. Lee, and S. Kettemann, 
 Phys. Rev. {\bf B 84}, 075453 (2011).
\bibitem{Liu2}
J. F. Liu, B. K. S. Chan and  J.  Wang, 
Nanotechnology, {\bf 23}(9):095201 (2012).


\bibitem{Mar} D. Marchenko, A. Varykhalov, M.R. Scholz, G. Bihlmayer, E.I. Rashba, A. Rybkin, A.M. Shikin and O. Rader,
Nature Communications {\bf 3}, 1232 (2012).


\bibitem{Richter} Ming-Hao Liu, Jan Bundesmann, and Klaus Richter, 
Phys. Rev. {\bf B 85}, 085406 (2012).

\bibitem{Patra} A. K. Patra, S. Singh, B. Barin, Y. Lee, J.-H. Ahn, E. del Barco, E. R. Mucciolo, and B. \"Ozyilmaz,
Phys. Lett. {\bf 101}, 162407 (2012).

 
 \bibitem{Dulbak} B. Dulbak,  M. B. Martin, C. Deranlot, B. Servet, S. Xavier, R. Mattana, M. C. Berger, W. A. De Heer, F. Petroff, A.  Anane, P. Seneor,  and Albert Fert
 Nature Phys. {\bf 8}, 557 (2012).

\bibitem{Zomer} P. J. Zomer, M. H. D. Guimarães, N. Tombros, and B. J. van Wees, 
Phys. Rev. {\bf B 86}, 161416(R) (2012).


\bibitem{Zhang1} Q. Zhang, K. S. Chan, Z. Lin and J. F. Liu, 
Phys. Lett. {\bf A} 377, 632 (2013).

\bibitem{Lenz} L. Lenz, D. F. Urban and D. Bercioux,
The European Physical Journal {\bf B 86}, 502 (2013). 


\bibitem{Shakouri} K. Shakouri, M. R. Masir, A. Jellal, E. B. Choubabi, and F. M. Peeters, 
Phys. Rev. {\bf B 88}, 115408 (2013).

\bibitem{Tang} Z. Tang, E. Shikoh, H. Ago, K. Kawahara, Y. Ando, T. Shinjo, and M. Shiraishi, 
Phys. Rev. {\bf B 87}, 140401 (2013).

\bibitem{Jaros} W. Han, R. K. Kawakami, M. Gmitra and J. Fabian,
Nature Nanotechnology {\bf 9}, 794 (2014). 

\bibitem{Kochan}  D. Kochan, M. Gmitra, and J. Fabian, 
 Phys. Rev. Lett. {\bf 112}, 116602 (2014). 
 
 \bibitem{Tuan} D. V. Tuan, F. Ortmann, D. Soriano, S. O. Valenzuela, S. Roche, 
 Nature Phys. {\bf 10}, 857 (2014). 

\bibitem{Gmitra} M. Gmitra and J. Fabian, 
Phys. Rev. {\bf B 92}, 155403 (2015). 

\bibitem{Ferrari} A. C Ferrari {\it et al.},
Nanoscale {\bf 7}, 4598 (2015).

\bibitem{Roch} S. Roch {\it et al.}
 2D Materials, {\bf 2}, 030202 (2015).

\bibitem{Zhang}  H. Zhang, Z.  Ma and J. F. Liu, 
Scientific Reports {\bf 4}, 6464 (2014). 

\bibitem{Drogler} M. Dr\"ogler {\it et al.}, 
 Nano Lett. 16, 3533 (2016).

\bibitem{Avsar} A. Avsar {\it et. al.},
 NPG Asia Mater. {\bf 8}, 274 (2016).

\bibitem{Ingla} J. Ingla-Aynés, R. J. Meijerink, and B. J. van Wees, 
 Nano Lett. 16, 4825 (2016).

\bibitem{Medina} B. Berche, F. Mireles, and E. Medina, 
Condensed Matter Physics {\bf 20}, 13702 (2017).

\bibitem{Li} X. Li, Z. Wu and J. Liu, 
Scientific Reports {\bf 7}, 6526 (2017).

\bibitem{Lin} Xiaoyang Lin {\it et al.},
Phys. Rev. Appl. {\bf 8}, 034006 (2017).
 
\bibitem{Amir} A. M.  Afzal, K. H. Min, B. M. Ko and J. Eom, 
RSC Adv. {\bf 9}, 31797 (2019). 
  
\bibitem{Shifei} S. Qi, Y. Han, F. Xu, X.  Xu and Z. Qiao,
Phys. Rev. {\bf B 99}, 195439 (2019).

\bibitem{Rashba} E. I. Rashba, Sov. Phys. Solid State {\bf 2}, 1109 (1960);
Y. A. Bychkov and E. I. Rashba,
JETP Lett. {\bf 39}, 78 (1984); 
 H. A. Engel  E. I. Rashba  and B. I. Halperin, 
arXiv:cond-mat/0603306.

\bibitem{Niu} J. Shi, P.  Zhang, D. Xiao, and Q. Niu, 
Phys. Rev. Lett. {\bf 96}, 076604 (2006).


\bibitem{David}  D. D. Awschalom, D. Loss, and N. Samarth, eds., 
{\it Semiconductor Spintronics and Quantum Computation} (Springer, Berlin, 2002).
\bibitem{Spintronics} I. Zutic, J. Fabian, and S. Das Sarma, 
 Rev. Mod. Phys. {\bf 76}, 323 (2004).

\bibitem{Hatano} N. Hatano, R. Shirasaki, H. Nakamura, 
Phys. Rev. {\bf A} 75, 032107 2007.

\bibitem{Matityahu} S. Matityahu, Y. Utsumi,  A. Aharony, O. Entin-Wohlman and C. A. Balseiro, 
Phys. Rev. {\bf B 93}, 075407 (2016).

\bibitem{AB1} Y. Avishai and Y. B. Band,
Phys. Rev. {\bf B 95}, 104429 (2017). 

\end{thebibliography}
\end{document}